\documentclass[pra,twocolumn,amsmath,longbibliography,nofootinbib]{revtex4-1}

\usepackage{graphicx}
\usepackage{amsmath}
\usepackage[dvipsnames]{xcolor}
\usepackage{soul}
\usepackage[normalem]{ulem}

\begin{document}

\title{Complexes formed in collisions between \\ ultracold alkali-metal diatomic molecules and atoms}

\author{Matthew D. Frye}

\affiliation{Joint Quantum Centre (JQC) Durham-Newcastle, Department of
Chemistry, Durham University, South Road, Durham DH1 3LE, United Kingdom}

\author{Jeremy M. Hutson}
\affiliation{Joint Quantum Centre (JQC) Durham-Newcastle, Department of
Chemistry, Durham University, South Road, Durham DH1 3LE, United Kingdom}

\date{\today}

\begin{abstract}
We explore the properties of 3-atom complexes of alkali-metal diatomic molecules with alkali-metal
atoms, which may be formed in ultracold collisions. We estimate the densities of vibrational states
at the energy of atom-diatom collisions, and find values ranging from 2.2 to 350~K$^{-1}$. However,
this density does not account for electronic near-degeneracy or electron and nuclear spins. We
consider the fine and hyperfine structure expected for such complexes. The Fermi contact
interaction between electron and nuclear spins can cause spin exchange between atomic and molecular
spins. It can drive inelastic collisions, with resonances of three distinct types, each with a
characteristic width and peak height in the inelastic rate coefficient. Some of these resonances are
broad enough to overlap and produce a background loss rate that is approximately proportional to
the number of outgoing inelastic channels. Spin exchange can increase the density of
states from which laser-induced loss may occur.
\end{abstract}

\maketitle

\section{Introduction}

Ultracold polar molecules have many potential applications, ranging from precision measurement
\cite{Zelevinsky2008, Hudson2011, Salumbides2011, Salumbides2013, Schiller2014, ACME2014,
Hanneke2016, Cairncross2017, Borkowski2018, ACME2018, Borkowski2019}, quantum
simulation~\cite{Barnett2006, Micheli:2006, Buchler2007, Macia2012, Manmana2013, Gorshkov2013} and
quantum information processing~\cite{Demille2002, Yelin2006, Zhu2013, Herrera2014, Ni2018,
Sawant:qudit:2020, Hughes2020} to state-resolved chemistry~\cite{Krems2008, Bell2009,
Ospelkaus:react:2010, Dulieu2011, Balakrishnan2016, Hu2019}. A few molecules such as
SrF~\cite{Shuman2010, Barry2014, McCarron2015, Norrgard2016}, YO~\cite{Hummon2013},
CaF~\cite{Zhelyazkova2014, Truppe2017, Anderegg2018}, YbF~\cite{Lim2018} and
SrOH~\cite{Kozyryev2017} have nearly closed vibronic transitions suitable for laser cooling. In
addition, a variety of polar alkali-metal diatomic molecules have been produced by association of
pairs of ultracold atoms, usually by magnetoassociation, followed by coherent optical transfer to
the ground rovibronic state. The molecules produced in this way include KRb~\cite{Ni:KRb:2008,
Voges:NaK:2020}, Cs$_2$~\cite{Danzl:v73:2008, Danzl:ground:2010}, Rb$_2$~\cite{Lang:cruising:2008},
RbCs~\cite{Takekoshi:RbCs:2014, Molony:RbCs:2014}, NaK~\cite{Park:NaK:2015, Seesselberg2018b,
Yang:K_NaK:2019}, NaRb~\cite{Guo:NaRb:2016}, NaLi~\cite{Rvachov:2017} and
NaCs~\cite{Cairncross:2021}.

All the alkali-metal diatomic molecules produced so far have been found to undergo collisional loss
in optical traps \cite{Ospelkaus:react:2010, Takekoshi:RbCs:2014, Park:NaK:2015, Guo:NaRb:2016,
Gregory:RbCs-collisions:2019, Voges:NaK:2020}, even in cases where there is no energetically
allowed 2-body reaction. In most systems the loss rate coefficients approach the predictions of a
``universal loss" model \cite{Idziaszek:PRL:2010, Idziaszek:PRA:2010} in which every molecular pair
that reaches short range is lost from the trap. For RbCs, however, detailed loss measurements that
include temperature dependence \cite{Gregory:RbCs-collisions:2019} have been used to determine the
parameters of a non-universal model \cite{Frye:2015} in which there is partial reflection at short
range.

Mayle \emph{et al.}\ \cite{Mayle:2012, Mayle:2013} proposed that the observed trap loss is due to
``sticky collisions", in which an initial bimolecular collision forms a long-lived complex that
survives long enough to collide with a third molecule. They estimated the densities of states
$\rho$ for the 4-atom complexes at the energy of the colliding molecules, and used arguments based
on random-matrix theory to estimate the resulting mean lifetime $\tau$ of the complex.
For (KRb)$_2$ they obtained $\rho>3000\ \mu\textrm{K}^{-1}$ and $\tau>150$~ms. In subsequent work,
Christianen \emph{et al.}\ \cite{Christianen:density:2019} obtained improved estimates of $\rho$,
taking fuller account of angular momentum constraints and using a more accurate representation of
the potential energy surface. The corresponding lifetimes, $\tau=6\ \mu$s for (NaK)$_2$, are too
short for most complexes to collide with a third molecule at the experimental densities. Christianen
\emph{et al.}\ \cite{Christianen:laser:2019} proposed that the complexes are instead excited by the
trapping laser, and showed that this can occur fast enough to account for the observed trap loss.
This proposal is supported by experiments on collisions of RbCs
\cite{Gregory:RbCs-complex-lifetime:2020, Gregory:atom-molecule:2021} and $^{40}$KRb \cite{Liu:K2Rb2:2021}, though
recent experiments on Na$^{40}$K \cite{Bause:2021} and on Na$^{39}$K and Na$^{87}$Rb
\cite{Gersema:2021} suggest that the complexes have longer lifetimes than
predicted in the absence of the trapping laser.

In parallel with the work on molecule-molecule collisions, experiments have been carried out on
atom-molecule collisions. The systems studied experimentally include $^{40}$K$^{87}$Rb with
$^{40}$K and Rb \cite{Ospelkaus:react:2010, Nichols:long-lived:2021},
$^{87}$RbCs with $^{87}$Rb and Cs \cite{Gregory:atom-molecule:2021},
Na$^{39}$K with Na and $^{39}$K \cite{Voges:NaK:2020, Voges:2021} and Na$^{40}$K with $^{40}$K \cite{Yang:K_NaK:2019, Wang:K_NaK:2021}. For each molecule, reaction is energetically allowed in a collision with the
lighter atom but forbidden in a collision with the heavier one. Fast collisional loss has been
observed in all cases where the reaction is energetically allowed.

For non-reactive atom-molecule systems, the picture is more complicated. Experiments have been carried out on a number of systems including $^{40}$K$^{87}$Rb with Rb \cite{Ospelkaus:react:2010, Nichols:long-lived:2021}, $^{87}$RbCs with Cs \cite{Gregory:atom-molecule:2021}, Na$^{39}$K with $^{39}$K \cite{Voges:NaK:2020, Voges:2021} and Na$^{40}$K with $^{40}$K \cite{Yang:K_NaK:2019}. For the last of these, Yang \emph{et al.}\ \cite{Yang:K_NaK:2019} observed narrow Feshbach resonances as a function of magnetic field; these resonances have been assigned as due to long-range states of triatomic complexes \cite{Wang:K_NaK:2021}.
A wide variety of behavior has been observed, ranging from near-universal loss to very slow loss, and no consistent picture is yet available. Recently, Nichols \emph{et al.}\ \cite{Nichols:long-lived:2021} directly probed complexes in collisions of $^{40}$K$^{87}$Rb with Rb; they measured lifetimes of 0.4 ms in the absence of the trapping laser, 5 orders of magnitude larger than theoretical predictions \cite{Christianen:density:2019}.

The purpose of the present paper is to explore the properties of the collision complexes that can be formed in collisions between alkali-metal diatomic molecules and atoms.
The structure of the paper is as follows. In section \ref{sec:ang-mom}, we consider
the angular momenta that are present in alkali-metal triatomic systems, and basic aspects of the
coupling between them. In section \ref{sec:lifetime}, we consider the relationship between
densities of states, lifetimes of complexes, and loss rates, including threshold effects. In
section \ref{sec:density}, we estimate the densities of vibrational states for short-range 3-atom
complexes near the atom-diatom collision threshold. In section \ref{sec:orb} we consider the
electronic structure of the complexes and the effect of orbital near-degeneracy. In section
\ref{sec:Fermi}, we present a model for the Fermi contact interaction between electron and nuclear spins, and show that it can drive spin exchange between atomic and molecular spins. In section \ref{sec:spin-exchange}, we consider the effect of spin exchange in atom-molecule collisions. We show that it can cause Feshbach
resonances of three distinct types, each of which produces peaks in the loss rate coefficient with characteristic widths and peak heights. Some of these resonances are broad enough to overlap and produce a background loss rate that is approximately proportional to the number of outgoing inelastic channels.
Finally, section \ref{sec:conclusions} presents perspectives and conclusions of the work.

\section{Angular momentum coupling \label{sec:ang-mom}}

There are 6 sources of angular momentum in a triatomic system AB+C formed from a singlet
molecule and an alkali-metal atom: the electron spin $S=1/2$, three nuclear spins $i_\textrm{A}$,
$i_\textrm{B}$, $i_\textrm{C}$, the diatomic rotation $n$ and the partial-wave quantum number $L$
for rotation of AB and C about one another.

An atom-diatom system in a single electronic state is governed by a 3-dimensional potential energy
surface $V(R,r,\theta)$. This is written here in Jacobi coordinates where $r$ is the diatom bond
length and $R$ and $\theta$ are the atom-diatom distance and angle. For the alkali-metal systems of
interest here, $V(R,r,\theta)$ is deep (of order 50~THz) and provides strong coupling between the
vibrational and rotational states of the diatomic molecule. Nevertheless, it is diagonal in the
total spin-free angular momentum $N$, which is the resultant of $n$ and $L$.

An alkali-metal atom C in a $^2$S state, with electron spin $S=1/2$ and nuclear spin
$i_\textrm{C}$, is characterized in zero field by its total spin
$f_\textrm{C}=i_\textrm{C}\pm\frac{1}{2}$. The two hyperfine states are separated by the hyperfine
splitting $(i_\textrm{C}+\frac{1}{2}) \zeta_\textrm{C}$, of order 1~GHz, where $\zeta_\textrm{C}$
is the scalar hyperfine coupling constant that arises from the Fermi contact interaction. In a
magnetic field, each hyperfine state is split into $2f_\textrm{C}+1$ Zeeman states labeled by
$m_{f,\textrm{C}}$. As will be seen below, the Fermi contact interaction in a triatomic complex can depend strongly on
geometry and provides a coupling that can be off-diagonal in $f_\textrm{C}$ and/or
$m_{f,\textrm{C}}$, while conserving the total spin projection
$m_{f,\textrm{tot}}=m_\textrm{A}+m_\textrm{B}+m_\textrm{C}+M_S$.

Our overall picture of the states of the triatomic complex is that the electronic interaction
potential creates a strongly coupled and potentially chaotic manifold of states for each spin
combination $(m_\textrm{A},m_\textrm{B},f_\textrm{C},m_{f,\textrm{C}})$, which correlates at long
range with an atom in state $(f_\textrm{C},m_{f,\textrm{C}})$ and a molecule in state
$(m_\textrm{A},m_\textrm{B})$. There is a weaker coupling between manifolds of the same
$m_{f,\textrm{tot}}$, due to the Fermi contact interaction, that may cause inelastic loss when
there are suitable open channels.

Both the electronic interaction potential and the Fermi contact interaction conserve the total
spin-free angular momentum $N$ and its projection $M_N$. There are weaker interactions arising from Zeeman, spin-rotation, and additional hyperfine interactions, some of which are off-diagonal in $N$ and $M_N$. These may play a role in sharp Feshbach resonances, but are unlikely to be
strong enough to influence background loss.

Atom-molecule pairs that reach short range may be excited by the trapping laser, producing laser-induced loss analogous to that observed in molecule-molecule collisions. This is possible both for direct collisions and for collisions that form long-lived complex; laser-induced loss is not in itself evidence of complex formation.

\section{Resonance widths and lifetimes of collision complexes \label{sec:lifetime}}

A quasibound state is often thought of as characterised by a width $\Gamma$ and a corresponding lifetime $\tau=\hbar/\Gamma$. However, these quantities need careful definition.
When a bound state is embedded in a scattering continuum, it produces a resonance whose width $\Gamma$ is governed by the matrix elements between the bound state and the continuum. If the state is well above the threshold of a single open channel, the scattering phase shift $\delta(E)$ follows the Breit-Wigner form,
\begin{equation}
\delta(E)=\delta_\textrm{bg} + \arctan\left(\frac{\Gamma}{2(E_0-E)}\right),
\label{eq:bw}
\end{equation}
where $\delta_\textrm{bg}$ is the background phase shift and $E_0$ is the resonance energy. If there are several open channels $a$, the S-matrix eigenphase sum \cite{Ashton:1983} follows the form \eqref{eq:bw} and $\Gamma$ is a sum of partial widths $\Gamma_a$ to the individual open channels.

When a quasibound state is probed by absorption spectroscopy from a single initial state, and the continuum itself is dark, the spectrum has a Lorentzian lineshape with width $\Gamma$ \cite{Feshbach:1958, Feshbach:1962, Fano:1961}. Conversely, excitation of the entire Lorentzian, by a laser that is broad compared to $\Gamma$, produces a non-stationary state (wavepacket) that decays into the continuum with lifetime $\tau=\hbar/\Gamma$ \cite{Bohm:2002}. Nevertheless, it is important to realize that, even when a bound state is spread out over a continuum, solutions of the time-independent Schr\"odinger equation still exist at each energy. Such solutions represent stationary states whose densities do not evolve in time. The rate at which a wavepacket can evolve is ultimately limited by its spread in energy, which can be much smaller than $\Gamma$. This is particularly important in ultracold systems, which may possess a very small energy spread characterized by their temperature. Complexes formed in ultracold systems may thus exhibit lifetimes limited by the temperature, which may be much longer than implied by the width of the underlying state.

\begin{figure}[tb]
\includegraphics[width=0.99\linewidth]{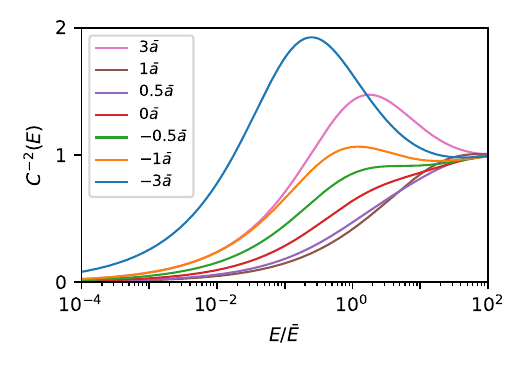}
\caption{QDT parameter $C^{-2}(E)$ as a function of reduced energy $E/\bar{E}$ for a selection of background scattering lengths $a_\textrm{bg}$. \label{fig:C-function}}
\end{figure}

The width of a Feshbach resonance is energy-dependent when the state lies just above threshold,
even if the underlying couplings are independent of energy. In quantum defect theory (QDT),
the partial width for decay of a resonance to an open channel $a$ is
\begin{equation}
\Gamma_a=\Gamma^{\textrm{s}}_a C^{-2}(E_a), \label{eq:narrowing}
\end{equation}
where $\Gamma^{\textrm{s}}_a$ is the short-range width and $C^{-2}(E_a)$ is a QDT function that
depends on the kinetic energy $E_a$ for channel $a$ \cite{Mies:1984, Mies:1984a, Mies:MQDT:2000,
Raoult:2004}. For an interaction potential $-C_6R^{-6}$, $C^{-2}(E_a)$ depends on the background
scattering length $a_\textrm{bg}$ and is a universal function when written in terms of the mean
scattering length $\bar{a}=(2\mu C_6/\hbar^2)^{1/4}\times 0.4779888\dots$ \cite{Gribakin:1993} and
the corresponding energy $\bar{E}=\hbar^2/(2\mu\bar{a}^2)$. Examples of $C^{-2}(E_a)$ are shown for
a variety of values of $a_\textrm{bg}/\bar{a}$ in Fig.\ \ref{fig:C-function}; it is proportional to
$E_a^{1/2}$ at limitingly low energy and in all cases it approaches 1 when $E_a\gg\bar{E}$. Because of this, $\Gamma_a \ll \Gamma_a^\textrm{s}$ near threshold.

For molecules with regular patterns of energy levels, such as most low-lying vibrational states,
the widths are unrelated to the spacings between levels and are often much smaller than them.
However, Mayle \emph{et al.}\ \cite{Mayle:2012, Mayle:2013} suggested that the vibrational states
of an atom-molecule or molecule-molecule collision complex are chaotic in nature, and may be
approximated by random-matrix theory (RMT).
Under these circumstances the mean short-range width of the states may be
written in terms of their mean spacing $d$ \cite{Mitchell:2010},
\begin{equation}
\bar{\Gamma}^\textrm{s}_a = \frac{T^\textrm{s}_a d}{2\pi}, \label{eq:RMTwidth}
\end{equation}
where $T^\textrm{s}_a$ is a transmission coefficient between 0 and 1 that governs the likelihood of
complex formation or decay for collisions that reach short range in channel $a$. Neglecting threshold effects, it is related to the unitarity deficit of the mean S matrix $\bar{S}$ by $T^\textrm{s}_a=1-|\bar{S}_{aa}|^2$
\cite{Mitchell:2010}.

Mayle \emph{et al.} \cite{Mayle:2012, Mayle:2013} gave estimates of lifetimes for collision
complexes in atom-molecule and molecule-molecule collisions based on $\bar{\tau}=\hbar/\bar{\Gamma}$ and
further estimated $\bar{\Gamma}$ as $d/(2\pi)$ to obtain the mean lifetime as $\bar{\tau}=2\pi\hbar\rho$. Christianen \emph{et al.}\ \cite{Christianen:density:2019}
obtained better estimates for $\rho$ but used the same procedure to estimate $\bar{\tau}$. However, this
neglects the reduction of $\Gamma$ both by threshold effects and by the short-range transmission
coefficient $T^\textrm{s}_a$. The actual widths of the resonances are thus likely to be
considerably smaller than the estimates of Mayle \emph{et al.}, and the lifetimes of the
complexes, once formed, are likely to be considerably larger than $\tau=2\pi\hbar\rho$.
In general, if the temperature $T$ is high enough to average over many resonances, $k_\textrm{B}T\gg
d$, then $\bar{\tau}=2\pi\hbar\rho$ is still the correct mean collisional time delay \cite{Frye:time-delay:2019}, but only a fraction of
collisions form complexes and those that do have extended lifetimes.

The effects of thresholds and of $T^\textrm{s}_a$ can be combined into a single transmission
coefficient $T_a$, which is generally not a simple product of the two contributions. In the
presence of partial reflection at short range, $T_a$ can be calculated from a non-universal QDT
theory \cite{Idziaszek:PRL:2010, Frye:2015}.
The formation of complexes, averaged over many resonances, can be described in the same way as
inelastic loss \cite{Croft:unified:2020} and characterized by a short-range parameter analogous to the loss parameter $y$ of Idziaszek and Julienne \cite{Idziaszek:PRL:2010}, where $T^\textrm{s}_a=4y/(1+y^2)$ \cite{Mitchell:2010}.
The collision complex may subsequently decay back to the incoming channel, or be lost by a secondary process such as collision with a 3rd body, laser excitation, or inelastic decay. If the system behaves chaotically, the mean width for decay back to the incoming channel is $\bar{\Gamma}_\textrm{inc} = T_\textrm{inc} d/(2\pi)$.
In the absence of other loss processes, this implies a mean lifetime $\bar{\tau}=2\pi\hbar\rho/T_\textrm{inc}$, but only if the thermal energy spread is large compared to $\bar{\Gamma}_\textrm{inc}$ as described above.
Christianen \emph{et al.}\ \cite{Christianen:lossy:2021} recently considered a lossy QDT model of such resonances. They concluded that if the complexes are lost rapidly once formed and $\bar{\Gamma}^\textrm{s}_\textrm{inc} = d/2\pi$, such that $T^\textrm{s}_\textrm{inc}=1$ in Eq.\ \eqref{eq:RMTwidth}, the loss rate is represented by $y=0.25$, as opposed to the $y=1$ implicit in the model of Croft \emph{et al.}\ \cite{Croft:unified:2020}.
The reasons for this apparent disagreement are not clear at present, but both methods rely on averaging over a large number of resonances.

The densities of states for atom-molecule collision complexes are much lower than those for
molecule-molecule complexes \cite{Christianen:density:2019}. In particular, as seen below, the mean
spacings between levels are far larger than both typical thermal energies and laser broadening. It is therefore not
appropriate to average across the resonance widths. Instead, we need to consider how the possible
presence of a single resonant state near threshold can enhance collisional loss.
In the following, we consider this
in terms of the resonant profile as the state crosses the incoming threshold, even if its energy
relative to threshold is almost fixed. As shown in the Appendix, the rate coefficient for resonant
inelastic loss due to a single resonance with no background loss, at limitingly low collision
energy, shows a Lorentzian peak of the form
\begin{equation}
k_2 = k_2^\textrm{univ} \left[1+\left(1-a_\textrm{bg}/\bar{a}\right)^2\right]
\frac{\Gamma^{\textrm{s}}_\textrm{inc}/\Gamma_\textrm{inel}}{[2(E-E_\textrm{res})/\Gamma_\textrm{inel}]^2+1}.
\label{eq:k2res}
\end{equation}
Here $k_2^\textrm{univ} = 4\pi\hbar\bar{a}/\mu$ is the universal rate coefficient at zero energy \cite{Idziaszek:PRL:2010}; values of $\bar{a}$ and $k_2^\textrm{univ}$ are given in Table \ref{tab:densities}. $\Gamma^{\textrm{s}}_\textrm{inc}$ is the short-range partial width for decay to the incoming channel, which may be well represented by Eq.\ \eqref{eq:RMTwidth}.
$\Gamma_\textrm{inel}$ is the sum of the partial widths for secondary loss processes, including both laser excitation and inelastic decay. The width of the peak is determined by $\Gamma_\textrm{inel}$, which is not subject to threshold effects in the incoming channel.
It is entirely possible for the resonantly enhanced loss
rate to exceed the universal rate, particularly when $\Gamma^{\textrm{s}}_\textrm{inc}>\Gamma_\textrm{inel}$.
Such supra-universal rates are observed, for example, near Feshbach resonances for Na$^{40}$K
with $^{40}$K \cite{Yang:K_NaK:2019}.

If $\Gamma_\textrm{inel}\gtrsim d$, which often occurs when there are many open channels,
multiple resonant features may overlap.
When only a few resonances overlap, the separate contributions may be approximately additive. However, when many resonances overlap, they reach the regime of Ericson fluctuations \cite{Mitchell:2010} and the total loss cannot be represented in this way. This situation has been considered by Christianen \emph{et al.}\ \cite{Christianen:lossy:2021}. Here we will consider only the case $\Gamma_\textrm{inel} < d$, which is applicable to typical alkali atom+diatom collisions.

\section{Densities of states for collision complexes \label{sec:density}}

The densities of states are much lower for 3-atom than for 4-atom complexes. Christianen \emph{et
al.}\ \cite{Christianen:density:2019} have developed a procedure for evaluating the density of
states, based on a semiclassical phase-space integral incorporating angular momentum constraints.
They obtain
\begin{align}
\rho(E)=& \frac{
g_{NJp} 4 \sqrt{2} \pi m_\textrm{A} m_\textrm{B} m_\textrm{C}
}{
h^3 (m_\textrm{A} + m_\textrm{B} + m_\textrm{C})
} \nonumber \\
& \times \int^{V<E} \frac{Rr}{\sqrt{\mu R^2+\mu_{\textrm{AB}}r^2}}[E-V(R,r,\theta)]^{1/2}
\nonumber\\ & R^2 \textrm{d}R\, r^2 \textrm{d}r\, \sin\theta \textrm{d}\theta. \label{eq:DOS}
\end{align}
Here $g_{NJp}$ is a parity factor that accounts for the absence of a conserved parity in classical
phase space, which is 1 for the systems considered here; $m_X$ is the mass of atom $X$;
$\mu_{\textrm{AB}}=m_\textrm{A}m_\textrm{B}/(m_\textrm{A} + m_\textrm{B})$ is the reduced mass of
the diatomic molecule AB; and $\mu=(m_\textrm{A} + m_\textrm{B})m_\textrm{C}/(m_\textrm{A} +
m_\textrm{B} + m_\textrm{C})$ is the atom-diatom reduced mass.
Christianen \emph{et al.}\ \cite{Christianen:density:2019} included a degeneracy factor
$g_{\textrm{ABC}}$ in this expression to account for equivalent nuclei, but here we take that into
account in considering hyperfine states below. The resulting density of vibrational states $\rho(E)$
is for a single electronic state and a single value of the spin-free total angular momentum $N$ and
its projection $M_N$.\footnote{We adopt the usual spectroscopic notation for a molecule with spin,
retaining $J$ for the angular momentum including electron spin and $F$ for the angular momentum
including electron and nuclear spins. We retain upper-case letters for quantum numbers of the
triatomic system, and use lower-case letters for quantum numbers that refer to individual colliding
species.} It also neglects fine and hyperfine structure.

Equation \eqref{eq:DOS} can be integrated numerically if a full interaction potential
$V(R,r,\theta)$ is available. However, if we make some approximations we can obtain analytic expressions. We wish to estimate the density of states that are strongly enough coupled to form a chaotic bath. Such states exist principally at short range. We therefore fix the first term in the integrand to its value at the equilibrium geometry, $R=R_\textrm{e}$ and $r=r_\textrm{e}$. We also approximate the potential to be isotropic and harmonic around the minimum,
\begin{equation}
V(R,r,\theta)=\frac{1}{2}k_r (r-r_\textrm{e})^2+\frac{1}{2}k_R (R-R_\textrm{e})^2. \label{eq:V_harm}
\end{equation}
The integral can now be evaluated analytically, giving
\begin{align}
\int [E-V(R,r,\theta)]^{1/2}
R^2 \textrm{d}R\, r^2 \textrm{d}r\, \sin\theta \textrm{d}\theta
=\frac{8\pi^2 E^{3/2}}{3\sqrt{k_rk_R}}. \label{eq:phase-approx}
\end{align}
Here the energy is relative to the potential minimum, and for the present purpose we evaluate $\rho$ at $E=D_\textrm{e}$, the energy of the triatomic minimum with respect to the energy of the separated atom and diatomic molecule. We therefore conclude that the density of short-range states around the atom-diatom threshold is likely to scale approximately as
\begin{equation}
\frac{m_\textrm{A} m_\textrm{B} m_\textrm{C}
}{
(m_\textrm{A} + m_\textrm{B} + m_\textrm{C})
}
\frac{R_\textrm{e}r_\textrm{e}}{\sqrt{\mu R_\textrm{e}^2+\mu_{\textrm{AB}}r_\textrm{e}^2}}
\frac{D_\textrm{e}^{3/2}}{\sqrt{k_rk_R}}. \label{eq:scale}
\end{equation}

The equilibrium geometries and binding energies $D_\textrm{e}$ for all the alkali-metal 3-atom
systems containing 2 or 3 identical atoms have been obtained from electronic structure calculations
\cite{Zuchowski:trimers:2010}. For each system, we estimate the force constant $k_R$ for the
atom-diatom vibrations assuming a Lennard-Jones potential in $R$. This gives
$k_R=72D_\textrm{e}/R_\textrm{e}^2$, where $D_\textrm{e}$ is the binding energy of the trimer with respect
to the atom+diatom threshold and $R_\textrm{e}=(C_6/D_\textrm{e})^{1/6}$; for $XY+X$ systems we use
$C_6$ coefficients from \cite{Zuchowski:vdW:2013} and for $X+X_2$ systems we use twice the $C_6$
coefficient for $X+X$ \cite{Derevianko:2010}. The force constant $k_r$ is taken to be the same as
that determined for the free diatomic molecule $XY$ from electronic spectroscopy
\cite{Linton:1999, Amiot:2002, Pashov:NaRb:2005, Coxon:2006, Docenko:2006, Staanum:2007, Pashov:2007, Gerdes:2008, Falke:2008, Tiemann:2009, Strauss:2010, Ivanova:LiRb:2011, Knoop:2011, Docenko:RbCs:2011, Takekoshi:RbCs:2012, Steinke:2012, Schuster:2012, Wang:NaRb:2013, Repp:2013, Ferber:2013, Julienne:Li67:2014, Maier:LiRb-Efimov:2015, Zhu:NaK:2017, Groebner:KCs:2017, Sovkov:2017, Hartmann:2019, Tiemann:2020, Guo:2021}.
The resulting densities of short-range states $\rho$ and mean level spacings $d$ for the
systems considered here are given in Table \ref{tab:densities}. We have selected one representative isotope of each element, and the small variations due to isotopic substitution can be calculated from Eq.\ \ref{eq:scale} if needed. The values range from 60.5 MHz for Cs$_2$+Cs to 9.48 GHz for Li$_2$+Li. Most of the differences come from the atomic and reduced masses, which vary by up to factor of 20 between systems. The value obtained by this method for K$_2$+Rb is within 25\% of that obtained by Christianen \emph{et al.}\ \cite{Christianen:density:2019} by evaluating Eq.\ \ref{eq:DOS} using a different model potential.

\begin{table*}[tbp]
\caption{Properties of atom+diatom systems: mean scattering length $\bar{a}$ and zero-temperature universal rate constant $k_2^\textrm{univ}$, together with densities of short-range states $\rho$ and mean spacings $d$ calculated from Eqs.\ \eqref{eq:DOS} and \eqref{eq:phase-approx}.\label{tab:densities}} \centering
\begin{ruledtabular}
\begin{tabular}{lccccc}
System 	& $\bar{a}$ (\AA)	& $\bar{E}/h$ (MHz)	& $k_2^\textrm{univ}$ (cm$^3$ s$^{-1}$)	& $k_\textrm{B}\rho(D_\textrm{e})$ ($\textrm{K}^{-1}$) 	& $d/h$ (MHz)	\\ \hline
 $^{7}$Li$^{7}$Li+$^{7}$Li &   21 &  245 & 3.58$\times 10^{-10}$ &  2.2 & 9480 \\
 $^{7}$Li$^{23}$Na+$^{7}$Li & 20.8 &  205 & 2.92$\times 10^{-10}$ & 3.87 & 5386 \\
 $^{7}$Li$^{40}$K+$^{7}$Li & 23.2 &  154 & 3.03$\times 10^{-10}$ & 5.92 & 3520 \\
 $^{7}$Li$^{87}$Rb+$^{7}$Li & 24.2 &  132 & 2.96$\times 10^{-10}$ & 7.31 & 2852 \\
 $^{7}$Li$^{133}$Cs+$^{7}$Li & 25.4 &  117 & 3.04$\times 10^{-10}$ & 9.36 & 2225 \\
 $^{7}$Li$^{23}$Na+$^{23}$Na &   26 & 57.3 & 1.6$\times 10^{-10}$ & 4.61 & 4520 \\
 $^{23}$Na$^{23}$Na+$^{23}$Na &   29 & 39.1 & 1.51$\times 10^{-10}$ & 9.49 & 2195 \\
 $^{23}$Na$^{40}$K+$^{23}$Na & 30.8 & 31.6 & 1.46$\times 10^{-10}$ & 18.6 & 1122 \\
 $^{23}$Na$^{87}$Rb+$^{23}$Na & 32.5 & 25.1 & 1.37$\times 10^{-10}$ & 26.7 &  779 \\
 $^{23}$Na$^{133}$Cs+$^{23}$Na & 34.3 & 21.4 & 1.37$\times 10^{-10}$ & 37.8 &  551 \\
 $^{7}$Li$^{40}$K+$^{40}$K & 36.2 & 17.8 & 1.34$\times 10^{-10}$ & 8.17 & 2549 \\
 $^{23}$Na$^{40}$K+$^{40}$K &   38 & 14.3 & 1.24$\times 10^{-10}$ & 20.6 & 1011 \\
 $^{40}$K$^{40}$K+$^{40}$K &   42 & 10.8 & 1.26$\times 10^{-10}$ & 37.3 &  559 \\
 $^{40}$K$^{87}$Rb+$^{40}$K & 42.9 & 9.05 & 1.13$\times 10^{-10}$ & 56.6 &  368 \\
 $^{40}$K$^{133}$Cs+$^{40}$K &   45 & 7.69 & 1.11$\times 10^{-10}$ & 77.9 &  267 \\
 $^{7}$Li$^{87}$Rb+$^{87}$Rb & 45.7 & 5.36 & 8.08$\times 10^{-11}$ & 14.6 & 1423 \\
 $^{23}$Na$^{87}$Rb+$^{87}$Rb & 47.3 & 4.66 & 7.77$\times 10^{-11}$ & 41.8 &  499 \\
 $^{40}$K$^{87}$Rb+$^{87}$Rb & 50.1 & 3.91 & 7.75$\times 10^{-11}$ & 75.5 &  276 \\
 $^{87}$Rb$^{87}$Rb+$^{87}$Rb & 53.4 & 3.06 & 7.35$\times 10^{-11}$ &  134 &  155 \\
 $^{87}$Rb$^{133}$Cs+$^{87}$Rb &   55 & 2.68 & 7.05$\times 10^{-11}$ &  200 &  104 \\
 $^{7}$Li$^{133}$Cs+$^{133}$Cs & 55.7 & 2.39 & 6.52$\times 10^{-11}$ & 26.4 &  790 \\
 $^{23}$Na$^{133}$Cs+$^{133}$Cs &   57 & 2.17 & 6.34$\times 10^{-11}$ & 76.2 &  273 \\
 $^{40}$K$^{133}$Cs+$^{133}$Cs & 59.6 & 1.89 & 6.33$\times 10^{-11}$ &  129 &  161 \\
 $^{87}$Rb$^{133}$Cs+$^{133}$Cs & 61.9 & 1.59 & 5.96$\times 10^{-11}$ &  240 & 86.7 \\
 $^{133}$Cs$^{133}$Cs+$^{133}$Cs & 65.2 & 1.34 & 5.88$\times 10^{-11}$ &  344 & 60.5
\end{tabular}
\end{ruledtabular}
\end{table*}

\section{Orbital near-degeneracy \label{sec:orb}}

To understand the chemical bonding in the collision complexes, it is useful first to consider the
homonuclear alkali-metal triatomic molecules \cite{Martins:1983}. At a geometry corresponding to an
equilateral triangle (point group $D_{3h}$), these systems are orbitally degenerate, with a single
electron in an orbital of symmetry $\textrm{e}'$. The resulting state has symmetry $^2$E, so is
subject to a Jahn-Teller distortion; the actual equilibrium geometry is an obtuse isosceles
triangle (point group $C_{2v}$). At this geometry the ground state has $^2$B$_2$ symmetry, but
there is a low-lying excited state of $^2$A$_1$ symmetry. At lower-symmetry geometries the two states are mixed. The two
resulting surfaces intersect along seams of conical intersections that include one at equilateral
geometries. The three equivalent potential minima on the lower surface are connected by low-energy
pathways through scalene and acute isosceles geometries; motion along these pathways gives rise to
the phenomenon known as pseudorotation, with characteristic energy-level patterns.

For heteronuclear 3-atom systems $X_2Y$, the situation is more complicated \cite{Zuchowski:trimers:2010}.
Some systems have ground states of $^2$B$_2$ symmetry with minima at isosceles geometries, while
others have ground states of $^2$A$'$ symmetry at scalene geometries (point group $C_s$).
Nevertheless, the principle remains that there are two electronic states of similar well depth that
cross and avoided-cross as a function of nuclear coordinates. The resulting short-range states are strongly coupled to one another, so the near-degeneracy of the partially filled orbitals produces almost a
doubling in the densities of states from the values of Section \ref{sec:density}.

\section{The Fermi contact interaction \label{sec:Fermi}}

The strongest hyperfine term for both the free atom and the triatomic collision complex is the
Fermi contact interaction. At long range this couples $S$ to $i_\textrm{C}$ for the free atom to
form $f_\textrm{C}$. At short range, however, all three nuclei experience significant spin
densities and $S$ couples to all of $i_\textrm{A}$, $i_\textrm{B}$ and $i_\textrm{C}$ to give
resultant $f_\textrm{tot}$. The coupling is of the form
\begin{equation}
\sum_{X=\textrm{A,B,C}} \zeta_X(R,\theta,\phi) \hat{\boldsymbol{i}}_X\cdot\hat{\boldsymbol{S}}, \label{eq:fermi}
\end{equation}
where $\hat{\boldsymbol{S}}$ and $\hat{\boldsymbol{i}}_X$ are the operators for the electron and
nuclear spin angular momenta. The coupling coefficients $\zeta_X$ are proportional to the product
of the corresponding nuclear magnetic moment and the electron spin density at nucleus $X$.
The spin densities are strongly dependent on geometry; as any one nucleus is pulled away from the
other two, the electron spin localizes on the separating atom, until the full hyperfine coupling of
the free atom is achieved as $R\rightarrow\infty$. However, in the strongly interacting region the
spin density is distributed between the three atoms and shifts substantially from one atom to
another as the complex vibrates. The spin density at the nucleus is sensitive only to the spin
population in atomic s orbitals, and is reduced if population is transferred to p orbitals.

There is some controversy over the Fermi contact interactions in alkali-metal triatomics near their
equilibrium geometries. For Na$_3$ \cite{Lindsay:1982} and K$_3$ \cite{Thompson:1981},
electron-spin resonance (ESR) studies of matrix-isolated species show that the s-orbital spin
densities on the 3 atoms sum to around 0.9 at the equilibrium geometry of the $^2$B$_2$ state, with
most of the density on the two equivalent atoms. For Li$_3$ \cite{Garland:1983} and Na$_2$Li
\cite{Mile:1995} the distribution is similar, but the densities sum to only 0.69 and 0.78,
respectively. The Li$_3$ result has been confirmed by a molecular-beam study \cite{Hishinuma:1992}.
These results accord with a physical picture in which relatively little spin density is
located in p orbitals. However, a molecular-beam microwave study on Na$_3$ \cite{Coudert:2002}
suggests that the Fermi contact interactions are much smaller than the ESR spin densities imply,
and this is supported by electronic structure calculations \cite{Hauser:2015}. These details do not
affect the basic physics discussed in the present paper: the important feature is that the spin
densities shift substantially between atoms as a function of wide-amplitude motions.

We have modeled the spin densities using a valence-bond method. This is related to the London-Eyring-Polanyi-Sato (LEPS) \cite{London:1929, Eyring:1930, Sato:LEPS:1955, Truhlar:1977} approach, which has been widely used for interaction potentials, including those of the alkali-metal triatomic molecules \cite{Varandas:1982, Varandas:1986}. In the LEPS approach, the two doublet surfaces are \cite{London:1929}
\begin{align}
&V^{\pm}(R_\textrm{AB},R_\textrm{BC},R_\textrm{AC})=J_\textrm{AB}+J_\textrm{BC}+J_\textrm{AC} \nonumber \\
&\pm
\left[\frac{1}{2}(K_\textrm{AB}-K_\textrm{BC})^2(K_\textrm{BC}-K_\textrm{AC})^2(K_\textrm{AC}-K_\textrm{AB})^2
\right]^{1/2}, \label{eq:LEPS}
\end{align}
where $R_{XY}$ is the separation of atoms $X$ and $Y$, and $J_{XY}$ and $K_{XY}$ are 2-atom Coulomb and exchange integrals. Values of these integrals can be estimated from the potential curves for the corresponding 2-atom systems by writing their singlet and triplet curves as
\begin{equation}
^{1,3}V_{XY}(R_{XY})=J_{XY}\pm K_{XY},
\end{equation}
with only the two-body singlet and triplet potentials \cite{Linton:1999, Amiot:2002, Pashov:NaRb:2005, Coxon:2006, Docenko:2006, Staanum:2007, Pashov:2007, Gerdes:2008, Falke:2008, Tiemann:2009, Strauss:2010, Ivanova:LiRb:2011, Knoop:2011, Docenko:RbCs:2011, Takekoshi:RbCs:2012, Steinke:2012, Schuster:2012, Wang:NaRb:2013, Repp:2013, Ferber:2013, Julienne:Li67:2014, Maier:LiRb-Efimov:2015, Zhu:NaK:2017, Groebner:KCs:2017, Sovkov:2017, Hartmann:2019, Tiemann:2020, Guo:2021} as input.

The London equation \eqref{eq:LEPS} \cite{London:1929} gives only the energies of the surfaces. However, Slater's derivation \cite{Slater:1931} of it allows calculation of the corresponding eigenfunctions and spin densities as well. The eigenfunctions are expressed as linear combinations of Slater determinants. Each determinant represents a configuration with one electron in the s orbital of each atom; for doublet states with $M_S=\frac{1}{2}$, two atoms have spin projection $m_s=\frac{1}{2}$ and one has projection $-\frac{1}{2}$. Remarkably, there are some atomic arrangements near isosceles geometries where one of the three configurations dominates. For such arrangements, the spin density on the isolated atom is actually dominated by $m_s=-\frac{1}{2}$, with the opposite sign to the overall spin. This is a result that is quite impossible for a spin-restricted Hartree-Fock (RHF) wavefunction with an unpaired electron in a single molecular orbital.

In a magnetic field, the Fermi contact interaction can mediate spin exchange between the electron
and nuclear spins (and indirectly among the nuclear spins) while conserving the total spin
projection $m_{f,\textrm{tot}}$. Each rovibrational state of the triatomic complex will split into $N_\textrm{hf}$ spin sublevels. If the atomic and molecular states are spin-stretched, $m_{f,\textrm{tot}}=\pm f_\textrm{max}$, where $f_\textrm{max}=S+i_\textrm{A}+i_\textrm{B}+i_\textrm{C}$, only a single sublevel exists for each rovibrational state of the complex. However, the number of sublevels
increases as $|m_{f,\textrm{tot}}|$ decreases. Neglecting exchange symmetry, there are $N_\textrm{hf}=4$ sublevels for $f_\textrm{max}-|m_{f,\textrm{tot}}|=1$; three of these have even exchange parity and one is odd. For larger deviations the numbers increase, and are easily evaluated, but depend on the
specific values of the spins.

\begin{figure}[tb]
\includegraphics[width=0.99\linewidth]{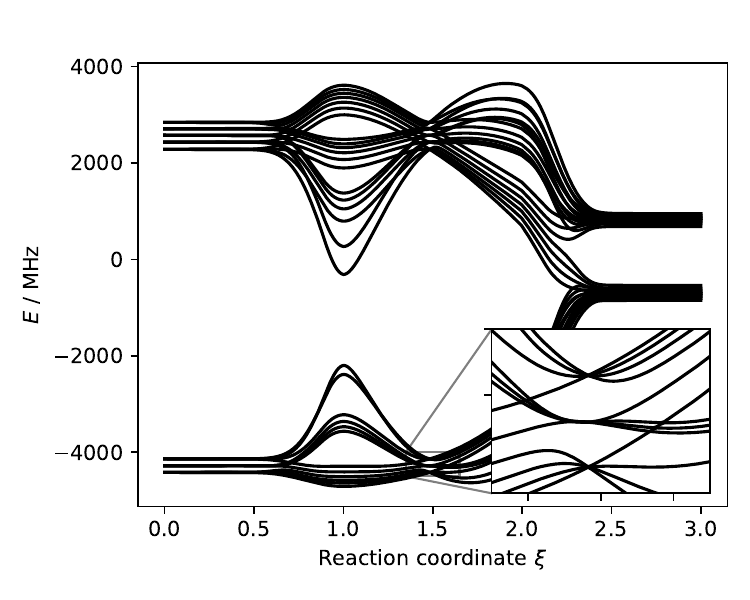}
\caption{Energies of hyperfine adiabats of $^{87}$Rb$_2{}^{40}$K with $m_{f,\textrm{tot}}=-3/2$ in a magnetic field of 200 G, with spin densities taken from the valence-bond model.
Along the coordinate $\xi$, the system starts from the separated Rb atom and KRb molecule ($\xi=0$); these approach each other with fixed Rb-K-Rb angle, to the absolute minimum ($\xi=1$), which is an obtuse isosceles triangle. The system then pseudorotates along the minimum-energy path to a local minimum at a scalene geometry ($\xi=2$), and finally dissociates by moving the K atom away from Rb$_2$, with fixed Rb-Rb-K bond angle ($\xi=3$). The inset shows an expansion of the adiabats around the midpoint of the pseudorotation, illustrating the presence of close avoided crossings that are likely to result in spin exchange.
} \label{fig:fermi}
\end{figure}

An example of the effect of shifting spin density is shown in Fig.\
\ref{fig:fermi}. This shows the Fermi contact contribution to the hyperfine energy for the reaction $^{87}$Rb +$^{40}$K$^{87}$Rb $\rightarrow$ $^{87}$Rb$_2$ + $^{40}$K along a selected reaction path that includes a rearrangement by pseudorotation. It is obtained by diagonalizing the Fermi-contact Hamiltonian (\ref{eq:fermi}) at each geometry, with the coupling coefficients $\zeta_X$ calculated from the valence-bond model. The resulting hyperfine adiabats are shown for $m_{f,\textrm{tot}}=-3/2$, which corresponds to the lowest hyperfine state of the reactants. In this case, $f_\textrm{max}-|m_{f,\textrm{tot}}|=6$ and there are $N_\textrm{hf}=32$ hyperfine sublevels.
At the center of the diagram, where
there is substantial exchange of spin density, the levels are spread out across an energy range
comparable to the atomic hyperfine splitting and their energies depend strongly on geometry.
There are many avoided crossings with a variety of widths, particularly along the pseudorotation section of the path, that are likely to result in spin exchange.

\section{Inelastic loss mediated by spin exchange}
\label{sec:spin-exchange}

There is a manifold of short-range vibrational states associated with each hyperfine sublevel. Each such manifold is likely to be chaotic in nature. Spin exchange allows a collision between atoms and molecules in one pair of states to access bound states in any manifold with the same $m_{f,\textrm{tot}}$. The density of states is enhanced to $N_\textrm{hf}\rho$, and all the states can cause Feshbach resonances.
However, these states are not necessarily equivalent, because the Fermi contact interaction that drives spin exchange is weaker than the anisotropic potential coupling. A given incoming channel couples strongly to one of these manifolds, but it may couple more weakly to the remaining $N_\textrm{hf}-1$ manifolds. This may be quantified with a spin-exchange parameter $z$, analogous to the isospin-mixing parameter used for reactions of compound nuclei \cite{Harney:1986},
\begin{equation}
z = \frac{4\pi^2 \overline{H_{ab}^2}}{d^2},
\end{equation}
where $H_{ab}$ is a matrix element of the coupling between states in manifolds $a$ and $b$. When $z\ll 1$, the mean partial width connecting the incoming channel to states in these other manifolds is $z\bar{\Gamma}_a$. Conversely, when $z\gg 1$, states in the $N_\textrm{hf}$ different manifolds are fully mixed and all contribute to the effective density of states that enters into RMT.

A full calculation of the spin-exchange parameter is beyond the scope of this work. It requires not only the mean level spacing $d$ but also the wavefunctions of the states in the two chaotic manifolds.
Such a calculation is near the limits of current theoretical methods, but may be achievable for lighter atom-diatom systems within a few years. Here we follow the practice in studies of compound nuclei \cite{Harney:1986}, and treat $z$ as a phenomenological parameter to be inferred from experimental results.

Spin exchange can also produce inelastic loss if there are channels lower in energy than the incoming state but with the same $m_{f,\textrm{tot}}$. We again take $^{40}$K$^{87}$Rb+$^{87}$Rb as our example \cite{Ospelkaus:react:2010, Nichols:long-lived:2021}. The lowest state is $(m_\textrm{K},m_\textrm{Rb},f_\textrm{Rb},m_{f,\textrm{Rb}})=(-4,3/2,1,1)$. Excitation of either nuclear spin of the molecule individually does not produce any inelastic channels that are accessible by spin exchange, but the combined excitation to $(m_\textrm{K},m_\textrm{Rb})=(-3,1/2)$ has the same $m_{f,\textrm{tot}}$ as the ground state, so spin exchange is allowed. However, the states are split by only the nuclear Zeeman effect, so the kinetic energy release is small, equivalent to 15~$\mu$K at a representative magnetic field of 200 G. The corresponding QDT function $C^{-2}(E)$ is typically around 0.1, so decay by this path is partially suppressed by threshold effects. If the atom is in an excited state, spin exchange is usually energetically allowed.
In such cases the kinetic energy releases are much larger, equivalent to 7 mK at 200 G, such that $C^{-2}(E)$ is close to 1, and loss is not suppressed by threshold effects.

Even when there are open channels that are energetically accessible, non-resonant spin exchange is likely to be slow. The avoided crossings in Fig. 2 occur deep in the potential well. Even for the widest such crossings, a Landau-Zener treatment estimates inelastic transition probabilities of the order of $10^{-3}$ for a double crossing in Rb+KRb. Significant loss is therefore expected only when there is resonant enhancement. If $z<1$, there are three general cases of resonant inelastic decay, which correspond to different values of $\Gamma^\textrm{s}_\textrm{inc}$ and $\Gamma_\textrm{inel}$ in Eq.\ \eqref{eq:k2res}. In the following, we relate the mean heights of resonant loss features to the universal rate, modified by the factor $[1+(1-a_\textrm{bg}/\bar{a})^2]$ in Eq.\ \ref{eq:k2res}, and their mean widths to $\bar{\Gamma}_a^\textrm{s}$ of Eq.\ \ref{eq:RMTwidth}.

The first case is where the resonant state is part of the manifold associated with the incoming channel and coupled to $N_\textrm{out}$ inelastic channels by spin exchange. We refer to the resulting resonances as incoming-manifold resonances. The coupling to the incoming channel is characterized by $\bar{\Gamma}^\textrm{s}_\textrm{inc}=\bar{\Gamma}^\textrm{s}_a$ and that to the inelastic channels by $\bar{\Gamma}_\textrm{inel}=z\bar{\Gamma}^\textrm{s}_a$. The height of the peak in the loss rate is thus multiplied by a factor $1/(zN_\textrm{out})$, but its width is divided by the same factor. When it is appropriate to average over many such resonances, their total contribution is thus independent of $z$, but in the atom+molecule case, where the density of states is too low for averaging to be appropriate, the narrowing decreases the probability of hitting such a resonance.

The second case is where the resonant state is part of a manifold associated with an open channel other than the incoming channel. We refer to these as outgoing-manifold resonances. In this case $\bar{\Gamma}^\textrm{s}_\textrm{inc}=z\bar{\Gamma}^\textrm{s}_a$ and $\bar{\Gamma}_\textrm{inel}=\bar{\Gamma}^\textrm{s}_a$. The height of the peak in loss rate is thus multiplied by a factor $z$. However the widths of the resonances are unaffected by $z$, so may be comparable to their spacings, according to Eq.\ \ref{eq:RMTwidth}. If there are $N_\textrm{out}$ open channels, each of them will support resonances of this type.

The third case is where the resonant state is part of a manifold associated with a closed channel. We refer to these as closed-manifold resonances. In this case $\bar{\Gamma}^\textrm{s}_\textrm{inc}=z\bar{\Gamma}^\textrm{s}_a$  and $\bar{\Gamma}_\textrm{inel} = zN_\textrm{out}\bar{\Gamma}^\textrm{s}_a$. The height of the peak in loss rate is thus reduced by a factor $N_\textrm{out}$ from the universal rate, and the width of the peak is multiplied by a factor $zN_\textrm{out}$.

\begin{figure}[tb]
\includegraphics[width=0.99\linewidth]{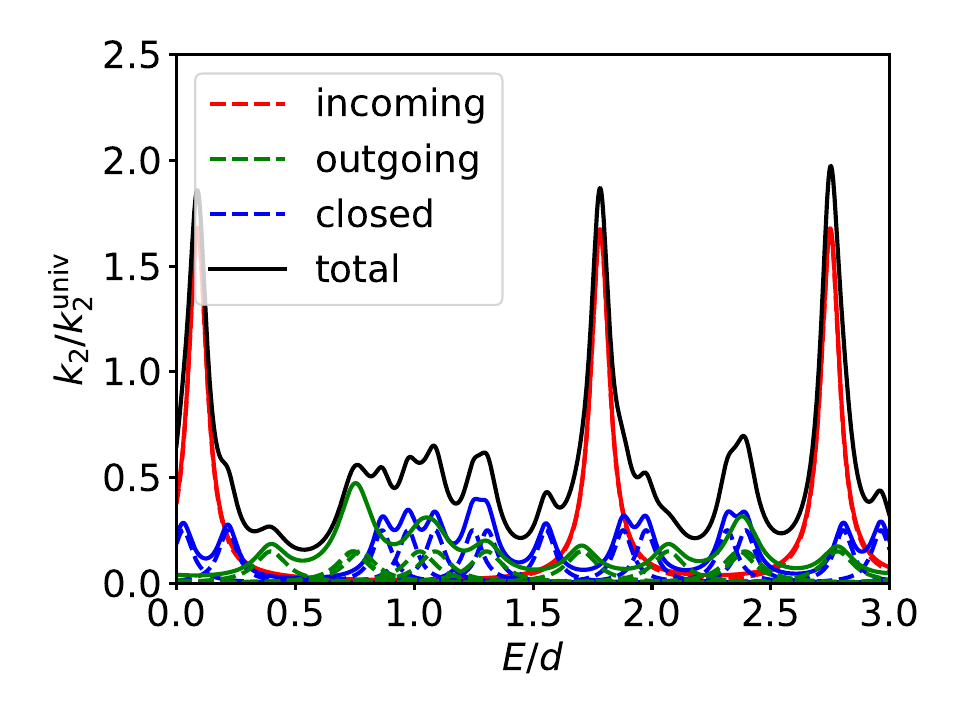}
\includegraphics[width=0.99\linewidth]{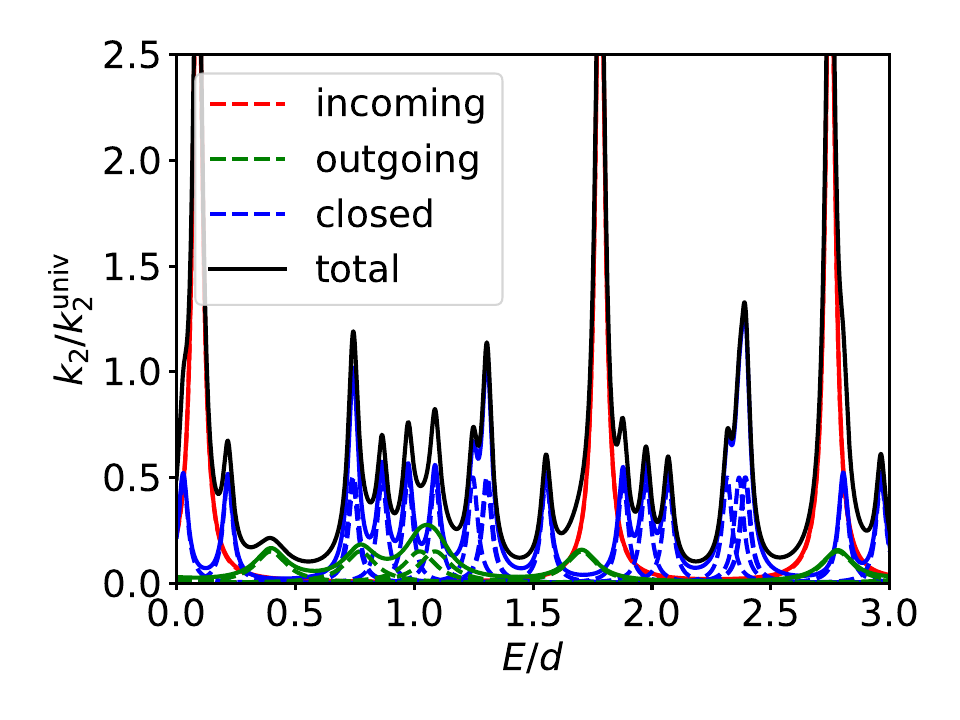}
\caption{Schematic examples of loss due to the three types resonance. Red, green and blue lines show incoming-, outgoing-, and closed-manifold resonances, respectively; dashed lines show individual resonant contributions from Eq.\ \ref{eq:k2res} and solid lines show simple sums neglecting interference effects. The coupling parameters used here are $T^\textrm{s}_a=1.0$ and $z=0.15$. We choose $N_\textrm{hf}=9$, corresponding to $f_\textrm{max}-|m_{f,\textrm{tot}}|=2$ for most alkali-metal trimer systems. The upper panel shows $N_\textrm{out}=4$ and the lower $N_\textrm{out}=2$.
\label{fig:cases}}
\end{figure}

Figure \ref{fig:cases} shows two schematic examples of the combined effect of these three types of resonance. For the purpose of illustration, we choose spin-exchange parameter $z=0.15$ and short-range transmission coefficient $T^\textrm{s}_a=1.0$. For each manifold, the bound states are randomly generated from a suitable chaotic distribution\footnote{We randomly generate matrices of dimension 20 from the Gaussian orthogonal ensemble, and unfold the resulting eigenvalues analytically using Wigner's semicircle law.} \cite{Mehta:rmt:1991, Guhr:1998} with mean spacing $d$.
The horizontal axis represents scanning the energy of the incoming channel across the pattern of short-range bound states. Figure \ref{fig:cases}(a) shows an example with $N_\textrm{hf}=9$ and $N_\textrm{out}=4$, so that within each interval $d$ there is 1 incoming-manifold resonance, 4 outgoing-manifold resonances, and 4 closed-manifold resonances. It may be seen that the outgoing-manifold and closed-manifold resonances make similar overall contributions to the inelastic rate, but the closed-manifold contribution is more structured because the underlying resonances are narrower. The incoming-manifold resonances give peaks that are high but relatively narrow (with width dependent on $T^\textrm{s}_a$), and usually will not overlap. Figure \ref{fig:cases}(b) shows an example with $N_\textrm{hf}=9$ and $N_\textrm{out}=2$, with the resonances in the same locations as before to facilitate comparison. Within each interval $d$ there are now 1, 2 and 7 incoming-, outgoing- and closed-manifold resonances, respectively. The major differences are that the peaks for incoming-manifold resonances and closed-manifold are higher and sharper, while the overall contribution of outgoing-channel resonances is reduced by the smaller $N_\textrm{out}$. An important feature that the two examples share is a weakly structured background loss from the outgoing-manifold resonances, whose mean height is
\begin{equation}
k_2^\textrm{bg} \approx \frac{z N_\textrm{out} T^\textrm{s}_a}{4}k_2^\textrm{univ},
\end{equation}
independent of $d$.

The loss rates in Fig.\ \ref{fig:cases} are shown as a function of the mean spacing $d$. In ultracold scattering, however, such rates are usually measured as a function of magnetic field, or some other external variable such as electric field. As a function of magnetic field $B$, channel threshold shift with respect to one another by $B\Delta\mu$, where $\Delta\mu$ is the difference in magnetic moments. For free alkali-metal atoms at low field, $\Delta\mu/h \approx 1.4/(i_\textrm{C}+\frac{1}{2})$ MHz/G for states with $m_{f,\textrm{C}}$ that differ by 1, and is typically between 0.7 MHz/G for atoms with $i_\textrm{C}=\frac{3}{2}$ ($^{23}$Na, $^{39}$K, $^{41}$K, $^{87}$Rb) and 0.3 MHz/G for $i_\textrm{C}=4$ ($^{40}$K). The magnetic moments of triatomic complexes are more complicated, but will be of similar magnitude. The horizontal axes in Fig.\ \ref{fig:cases} correspond to $3d$ and thus covers the equivalent of hundreds or even thousands of Gauss. The general conclusion is that atom-molecule resonances due to states that belong to the RMT bath are very broad indeed.

The largest body of experimental results for losses in alkali-metal atom-molecule systems is for Na$^{40}$K with $^{40}$K. Yang \emph{et al.}\ \cite{Yang:K_NaK:2019} measured loss rates as a function of magnetic field for more than 20 combinations of atomic and molecular states. Their main focus was on narrow Feshbach resonances, but they also measured loss rate coefficients in several windows of magnetic field, including near 90~G and near 102~G, shown in Figs.\ S2 and S3 of their Supplemental Material. In cases where no resonant features are visible, the background loss rates generally increase with $N_\textrm{out}$, reaching values between 0.5 and 1 times the universal loss rate when $N_\textrm{out}$ is 3 to 5. However, there are also a few state combinations that show unstructured loss at up to twice the universal rate. For Na$^{40}$K with $^{40}$K, $d=1033$~MHz and a typical value of $\Delta\mu/h$ is 0.3 MHz/G, so even the narrowest resonances in Fig.\ \ref{fig:cases}(b) have widths of order 200~G for $z=0.15$. Both the general increase in background loss rates with $N_\textrm{out}$ and the occasional state combinations with supra-universal rates are entirely consistent with the behaviour shown in Fig.\ \ref{fig:cases}.

In addition to inelastic loss to lower-lying open channels, there is the possibility of loss due to laser excitation of complexes by the trapping laser. This is likely to introduce loss from all channels accessible by spin exchange, with a contribution $\bar{\Gamma}_\textrm{laser}$ to $\bar{\Gamma}_\textrm{inel}$ that depends on the laser intensity but is independent of $\bar{\Gamma}_a^\textrm{s}$ and hence of the density of states. The effect of this depends on the relative strength of this decay to any existing inelastic loss. Nichols \emph{et al.}\ \cite{Nichols:long-lived:2021} have measured decay equivalent to $\bar{\Gamma}_\textrm{laser}/h\approx 5$ MHz for $^{40}$KRb+Rb at typical optical trap intensities, and it seems likely that other alkali-metal atom+molecule systems will be comparable. This value is substantially smaller than our estimates of $\bar{\Gamma}_a^\textrm{s}$ for any of the systems considered here. Laser-induced loss is thus likely to have little effect on loss due to any outgoing-manifold resonances, but it may reduce the height and increase the width of incoming-manifold or closed-manifold resonances if $\bar{\Gamma}_\textrm{laser}\gtrsim z\bar{\Gamma}_a^\textrm{s}$. More importantly, it may induce both incoming-manifold and closed-manifold resonances even in systems with no open inelastic channels, such as those involving atoms and molecules in their absolute ground states. However, the likelihood of encountering such a resonance for a particular incoming state at a single field is approximately $(N_\textrm{hf}-1)\bar{\Gamma}_\textrm{laser}/d$, which is usually fairly small.

\section{Conclusions}
\label{sec:conclusions}

We have developed the theory of the triatomic complexes that can be formed in ultracold collisions between alkali-metal diatomic molecules and atoms. We have estimated the densities of vibrational states of the complexes near the energy of the colliding particles, based on the properties of the diatomic molecules and the calculated binding energies of the trimers. The resulting densities range from 2.2 to 350~K$^{-1}$.
We have considered the angular momentum couplings present in the complexes and the resulting fine and hyperfine structure. The largest such term is the Fermi contact interaction between the electron and nuclear spins. We have presented a model of this interaction, based on valence-bond theory, and shown that it varies substantially with the geometry of the complex because the unpaired spin moves between the atoms as the complex vibrates.

Our overall picture is that each pair of atomic and molecular spin states is associated with a manifold of vibrational states of the triatomic complex. Each such manifold is likely to be chaotic in nature. The Fermi contact interaction can couple these manifolds, and can thus drive spin exchange between spins on the atom and the molecule. The degree of coupling between manifolds is characterized by a spin-exchange parameter $z$, which depends on the ratio between off-diagonal matrix elements of the Fermi contact interaction and the mean spacing between vibrational levels, $d$. This parameter may be substantial in the alkali-metal triatomic complexes.

We have developed the theory of resonant low-energy scattering in the presence of several chaotic manifolds that are weakly coupled to one another. We find that there can be Feshbach resonances of three distinct types, which we term incoming-, outgoing- and closed-manifold resonances. Each type of resonance can cause peaks in the rate coefficient for inelastic scattering with characteristic peak heights and widths that depend on $z$ and the number of outgoing inelastic channels. For atom-diatom systems, the resonances due to states in the chaotic bath are very broad compared to the spread of kinetic energies at ultracold temperatures, so it is not appropriate to average over their widths in ultracold atom-diatom collisions. Instead, the scattering properties at a specific collision threshold depend on where it is placed with respect to the essentially random pattern of individual bath states. Nevertheless, the resonances due to outgoing-manifold resonances are broad enough that they may overlap to produce a background loss for most incoming states, particularly when there are several outgoing inelastic channels that are accessible from the incoming state by spin exchange.

Resonances due to states in the chaotic bath are different from those arising from near-threshold states of triatomic complexes, as observed in Na$^{40}$K+$^{40}$K \cite{Wang:K_NaK:2021}. Such states spend most of their time in the long-range tail of the interaction potential; they are relatively weakly coupled to one another and to the incoming and outgoing scattering states, so they can produce much narrower resonances than the states considered here. They do not form part of the chaotic bath of states, so there is no reason to expect their widths to be related to their spacings.

Atom-diatom complexes have much lower densities of states than diatom-diatom complexes. Nevertheless, some features of the theory developed here may apply in the diatom-diatom case. In particular, the densities of states in diatom-diatom complexes \cite{Christianen:density:2019} can be 4 orders of magnitude larger than those obtained here for atom-diatom complexes. Since the mixing between manifolds is governed by the \emph{ratio} of couplings to level spacings, much smaller couplings are sufficient to cause mixing in the diatom-diatom case. Even very small interactions such as the nuclear electric quadrupole coupling \cite{Aldegunde:singlet:2017} might be large enough.

\begin{acknowledgments}
This work was supported by the U.K. Engineering and Physical Sciences Research Council (EPSRC)
Grant No.\ EP/P01058X/1.
\end{acknowledgments}

\appendix*
\section{Scattering length and loss rate across a decayed resonance}

Hutson \cite{Hutson:res:2007} has described the behavior of a Feshbach resonance close to
threshold in the presence of inelastic decay. The general expression for an S-matrix element in the
vicinity of a resonance is
\begin{equation}
S_{ab}(E)=S_{\textrm{bg},ab}-\frac{\textrm{i}g_{a}g_{b}}{E-E_\textrm{res}+\textrm{i}\Gamma/2}.
\end{equation}
Here, $E_\textrm{res}$ is the resonance energy, $g_{a}$
is in general complex and characterizes the partial width $\Gamma_a=|g_a^2|$, and
$\Gamma=\sum_a\Gamma_a$ is the total width. We are interested in the diagonal S-matrix element
in the incoming channel, $S_\textrm{inc}$;
we take the collision energy $E-E_\textrm{inc}$ to be small, such that we
can approximate the background S-matrix element to be $S_{\textrm{bg},\textrm{inc}}=1$; the more general case $|S_{\textrm{bg},\textrm{inc}}|=1$ gives the same results for inelastic loss due to resonances, but we choose this specific value for simplicity.
With this choice, the product $g_\textrm{inc}^2$ must be real and non-negative and can be replaced by
$\Gamma_\textrm{inc}$, the partial width to the incoming channel.
This gives
\begin{equation}
S_\textrm{inc}=1-\frac{\textrm{i}\Gamma_\textrm{inc}}{(E-E_\textrm{res})+\textrm{i}\Gamma/2}. \label{eq:S00}
\end{equation}
The total width is $\Gamma=\Gamma_\textrm{inc}+\Gamma_\textrm{inel}$, where $\Gamma_\textrm{inel}$
characterizes the decay to all loss channels, whether inelastic, reactive, or light-induced. The partial width in the incoming
channel is narrowed by threshold effects, $\Gamma_\textrm{inc}=\Gamma^{\textrm{s}}_\textrm{inc}C^{-2}(E-E_\textrm{inc})$, as described by
Eq.\ \ref{eq:narrowing}.

The QDT function $C^{-2}(E_\textrm{kin})$ can be calculated explicitly from either numerical or analytic
solutions \cite{Gao:C6:1998} with an appropriate asymptotic potential. Its leading term for an
asymptotic potential $-C_6/R^6$ is \cite{Mies:MQDT:2000, Raoult:2004, Julienne:2006}
\begin{equation}
C^{-2}(E_\textrm{kin})=k\bar{a}\left[1+\left(1-a_\textrm{bg}/\bar{a}\right)^2\right], \label{eq:C_func}
\end{equation}
where $k=\sqrt{2\mu E_\textrm{kin}}/\hbar$ is the wavevector and $a_\textrm{bg}$ is the background (non-resonant) scattering length.

We could now directly calculate inelastic cross sections and rate coefficients from Eq.\
\eqref{eq:S00}, but it is convenient first to consider the complex scattering length. This is
defined as \cite{Hutson:res:2007}
\begin{align}
a(k)&=\alpha-\textrm{i}\beta \nonumber \\
&=\frac{1}{\textrm{i}k}\frac{1-S_\textrm{inc}}{1+S_\textrm{inc}} \label{eq:a_def}
\end{align}
and the two-body loss rate coefficient is
\begin{equation}
k_2=\frac{4\pi\hbar\beta}{\mu(1+k^2|a|^2+2k\beta)}. \label{eq:k2def}
\end{equation}
This becomes constant and proportional to $\beta$ at low energy; the universal rate at zero energy, $k_2^\textrm{univ}$
corresponds to $\beta=\bar{a}$ \cite{Idziaszek:PRL:2010}. Substituting Eq.\ \eqref{eq:S00} into
Eq.\ \eqref{eq:a_def} and using Eq.\ \eqref{eq:C_func} gives
\begin{equation}
a=\bar{a}\left[1+\left(1-a_\textrm{bg}/\bar{a}\right)^2\right]
\frac{\Gamma^{\textrm{s}}_\textrm{inc}/\Gamma_\textrm{inel}}{2(E-E_\textrm{res})/\Gamma_\textrm{inel} + \textrm{i}},
\end{equation}
and so
\begin{equation}
\beta=\bar{a}\left[1+\left(1-a_\textrm{bg}/\bar{a}\right)^2\right]
\frac{\Gamma^{\textrm{s}}_\textrm{inc}/\Gamma_\textrm{inel}}{[2(E-E_\textrm{res})/\Gamma_\textrm{inel}]^2+1}.
\end{equation}
This can be converted to a rate through Eq.\ \ref{eq:k2def}. We make the simplifying assumption that $k|a|\ll 1$, such that the denominator in Eq.\ \ref{eq:k2def} reduces to just $\mu$, and write the result in terms of $k_2^\textrm{univ} = 4\pi\hbar\bar{a}/\mu$. This gives
\begin{equation}
k_2 = k_2^\textrm{univ} \left[1+\left(1-a_\textrm{bg}/\bar{a}\right)^2\right]
\frac{\Gamma^{\textrm{s}}_\textrm{inc}/\Gamma_\textrm{inel}}{[2(E-E_\textrm{res})/\Gamma_\textrm{inel}]^2+1}.
\end{equation}
which is Eq.\ \ref{eq:k2res}.
This is a Lorentzian peak with width determined only by $\Gamma_\textrm{inel}$, and a peak height
that reaches at least the universal rate if $\Gamma^{\textrm{s}}_\textrm{inc}\geq\Gamma_\textrm{inel}$.

\bibliographystyle{long_bib}
\bibliography{../all,Polarizability}

\end{document}